\documentstyle[epsf]{mn}
\def\fun#1#2{\lower3.6pt\vbox{\baselineskip0pt\lineskip.9pt
  \ialign{$\mathsurround=0pt#1\hfil##\hfil$\crcr#2\crcr\sim\crcr}}}
\def\lap{\mathrel{\mathpalette\fun <}}

\begin{document}
\title{Interpreting the Clustering of Radio Sources}

\author[C.~M.~Cress and M.~Kamionkowski]{Catherine M. Cress$^1$
and Marc Kamionkowski$^2$\\
$^1$Department of Astronomy, Columbia University, 538 West 120th
Street, New York, NY 10027~~U.S.A.\\
$^2$Department of Physics, Columbia University, 538 West 120th
Street, New York, NY 10027~~U.S.A.}

\maketitle
\begin{abstract}
We develop the formalism required to interpret, within a CDM 
framework, the angular clustering of sources in a deep radio 
survey. The effect of nonlinear evolution of density 
perturbations is discussed as is the effect of
the assumed redshift distribution of sources. We also investigate 
what redshift ranges contribute to the clustering signal at 
different angular scales. Application of the formalism is focussed on
the clustering detected in the {\sl FIRST} survey but measurements
made for other radio surveys are also investigated. We comment on the 
implications for the evolution of clustering.
\end{abstract}

\maketitle  

\section{INTRODUCTION}

The canonical cold dark matter (CDM)
model for the origin of structure in the Universe 
provides a concise qualitative description of
large-scale-structure data spanning several
orders of magnitude in distance scales, but on closer 
inspection, there seem to be anomalies.
For example, optical and infrared surveys seem to indicate more
power on large scales relative to that on small scales (Peacock
\& Dodds 1994). A number of variations on the standard CDM model have been
suggested which improve the agreement between some data and model predictions. 
These include (but are not limited to) the introduction of a
cosmological constant (Efstathiou, Sutherland \& Maddox 1990;
Kofman, Gnedin \& Bahcall 1993; Krauss \& Turner 1995; Ostriker
\& Steinhardt 1995) or a lower Hubble constant (Bartlett et
al. 1994).  However, a problem with discriminating between these
various models from large-scale-structure measurements is that
luminous matter may
be `biased' relative to the mass distribution (Kaiser 1984) and this bias
is likely to evolve with time and scale (see, for example, 
Matarrese et al. 1996). Further investigation of the bias
of luminous objects is thus important in the recovery of the primordial
power spectrum from large scale structure data.

In this paper, we investigate the implications of new data
obtained from deep radio surveys: in particular,
the VLA {\sl FIRST} (Faint Images of the Radio Sky at Twenty
centimeters) survey (Becker, White \& Helfand 1995; White et
al. 1996).

The goal of {\sl FIRST} is to survey the 10,000 $\rm deg^{2}$ scheduled
for inclusion in the Sloan Digital Sky Survey down to a 
flux-density threshold  of 1 mJy.
At present, the survey covers almost 3,000 $\rm deg^{2}$ of sky where 
$07^{h}16$ $^{<}_{\sim}$ $\alpha$ $^{<}_{\sim}$ $17^{h}40$ and 
$22^{\circ}$ $^{<}_{\sim}$ $\delta$ $^{<}_{\sim}$ $42^{\circ}$.
This yields a catalog of $\sim 250,000$ sources,
about one third of which are in double-lobed and multi-component sources. 
The survey has been shown to be 95\% complete down to 2 mJy and 80\%
complete to 1 mJy. 
The mean redshift of the radio sources in the survey 
is at $z\simeq 1$, as opposed to redshifts $z\lap 0.1$
characteristic of sources in optical or infrared surveys, so the
typical physical distances for fixed angular separations are
larger.  Therefore, clustering of {\sl FIRST} radio sources has the potential
to probe the power spectrum on large scales and at earlier epochs.

The first high-significance detection of an angular correlation function 
for radio sources was presented in Cress et al. (1996).
Using the CF-estimator proposed by Landy and Szalay (1993) on the first
1550 $\rm deg^{2}$ of the survey, it was found that the correlation
function between $0.02^{\circ}$ and $2^{\circ}$ is well-fit by a power 
law of the form $A\theta^{1-\gamma}$ where $A\approx 
2\times 10^{-3}$ and $\gamma\approx 2.2$. Results using a simpler
estimator were also presented but when the survey was later expanded to
include a new $\sim1500\,\rm deg^{2}$ it became evident that the LS
estimator was the more robust (Cress et al. 1997). Correlation-function 
measurements shown here are determined using the LS estimator in the 
expanded survey area. 

The purpose of the work presented here is to develop the formalism 
required to interpret, within a CDM framework,
angular-correlation-function measurements
of deep surveys and to apply this to the observational 
results of Cress et al. (1996). We also discuss applications to 
measurements made from other radio surveys.
We restrict our investigation to spatially-flat
models (allowing for the possibility of a non-zero
$\Lambda$).
Given a power spectrum, calculation of the angular correlation function is
straightforward and similar to that for optical or infrared
surveys.  However, there are two important issues which we
address here.  (1) We investigate the uncertainties in the
predictions which arise from imprecise knowledge of the survey
redshift distribution. (2) Since the typical redshifts of
sources in the {\sl FIRST} survey are of order unity, the
evolution of the power spectrum must be considered in the analysis
or else there may be errors of order unity in the deduced
primordial power spectrum. We include estimates of the non-linear evolution
of the power spectrum.
 
The plan of the paper is as follows:  In \S2, we discuss
the redshift distribution inferred for the sample.  In \S3, 
we discuss the time evolution of the power spectrum, and in
\S4, we discuss the calculation of the angular correlation
function. Results are presented and discussed in \S5.  
We make some concluding remarks in \S6.

\section{THE REDSHIFT DISTRIBUTION}
We use redshift-distribution estimates derived from two sources: (i) the
1.4-GHz radio luminosity function (RLF) presented by Condon (1984) and 
(ii) a set of 2.7-GHz LF estimates presented by Dunlop \& Peacock
(1990).  

In estimating the RLF, Condon used faint radio sources with 
optical counterparts
in the UGC catalog to determine a $local$ RLF for spiral galaxies. He 
combined this with Auriemma's (1977) estimate of the local
RLF for elliptical galaxies and then found a model for the RLF at higher
redshifts by allowing the local population to have evolved in density
and luminosty. The functions of redshift that describe the evolution
were constrained using number counts (down to sub-mJy thresholds) 
at 1.4\,GHz and at other frequencies, spectral-index measurements and 
redshift measurements for a few very bright sources.
 
Dunlop \& Peacock (1990) (DP) estimated the 2.7-GHz RLFs of 
steep- and
flat-spectrum sources separately using similar constraints as those
used by Condon as well as some new data which included redshifts 
estimated from K-band photometry. To 
demonstrate the possible errors in the RLFs resulting from these imprecise
redshift estimates, they analyze two distributions: one where assigned 
redshifts are the mean values for galaxies of a given K-band luminosity 
(mean-$z$ distribution) and one (the high-$z$ distribution)
where redshifts are
larger than the average (they give reasons why one might expect this kind of 
bias). Seven models were presented and parameters 
for all seven models were determined for both mean- and high-$z$
distributions. 

Here, we obtain the redshift distribution of
{\sl FIRST} sources by integrating the RLF's over all luminosities that produce
flux-densities greater than 1\,mJy. The DP models were first shifted to 
1.4\,GHz
assuming a spectral index of $\alpha=0.85$ for steep-spectrum sources and 
a spectral index of $\alpha=0$ for flat-spectrum sources 
(where the luminosity is related to the frequency by L$\propto\nu^{-\alpha}$).

Figure 1 shows the redshift distribution $dN/dz$ for Condon's LF
and for a selection of the models
given in DP for a flux-density limit of 1 mJy. Mean- and
high-$z$ distributions produce similar results and the
differences between DP's models 1 through 5 are smaller than the differences
shown in the Figure.  In Condon's model, the sharp increase in the number
of sources near $z=0$ is due to 
starbursting galaxies.  For larger flux-density limits, this
low-$z$ peak is reduced considerably, and we have checked that our
redshift distributions for larger flux-density limits agree with
those shown in Fig. 6 of Loan, Wall \& Lahav (1996).
\begin{figure}
\epsfxsize=85truemm 
\epsffile{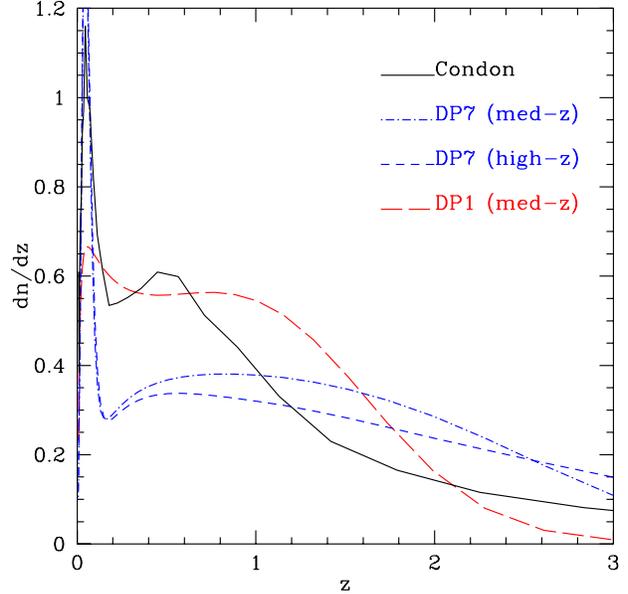}
\caption[figure1]{The redshift distribution $dN/dz$
derived from Condon's LF and from a selection of the models
given in DP for a flux-density limit of 1 mJy.}
\end{figure}

\section{THE EVOLVED POWER SPECTRUM}

In CDM models, one starts with a scale-free primordial power
spectrum, $P(k)\propto k^n$ (where $k$ is the comoving wavenumber). 
This $P(k)$ is then `processed' according a transfer function for CDM. 
At the epoch at which structure begins to grow, 
\begin{equation}
 P(k)=P_0\,k^n\,T^2(k),
\label{pkprim}
\end{equation}
where $T(k)$ for CDM is given by Bond \& Efstathiou (1984) and $P_0$ is 
the normalisation. We normalise to {\sl COBE} 4-year data using
Liddle et al.'s (1996) formula which is valid for spatially-flat models.
We will refer to this $processed$ spectrum as the primordial power spectrum.

In the linear regime, the time evolution of the power spectrum can be
calculated analytically and depends only on the expansion of the
Universe. It is given by
\begin{equation} 
P(k,z) = P(k,z=0)G^2[\Omega_0(z),\Omega_\Lambda(z)]/(1+z)^2, 
\label{linearevolution}
\end{equation}
where the growth factor, $G$, is given in Carroll, Press \& Turner 
(1992), and $G(\Omega_0=1, \Omega_\Lambda=0)=1$.

In the highly non-linear regime, the CF obtained from a scale-free primordial
spectrum obeys a simple scaling relation (Groth \& Peebles 1977).
One can interpolate between the linear and highly non-linear regimes to 
produce semi-analytic models for clustering in the quasi-linear
regime which can then be accurately fit to
results from $N$-body simulations (Hamilton 1991; Jain, Mo, \& White 1995; 
Peacock \& Dodds 1996). PD use the dimensionless power spectrum 
$\Delta^2(k,z)\equiv (2\pi^2)^{-1} k^3 P(k,z)$. In our notation, 
the non-linear power spectrum is then given by 
\begin{equation}
     P_{NL}(k,z)={k^3_{NL}\over 2\pi^2}f_{NL}[{2\pi^2\over k^3_{L}}P(k,z)], 
\label{pnlk}
\end{equation}
where the linear and nonlinear scales are related by
$k_L=[1+\Delta_{NL}^2(k_{NL})]^{-1/3} k_{NL}$.  The form of the
function, $f_{NL}$, is given in PD for various values of
$\Omega_0$.  We take into account the time evolution of the
power spectrum by evaluating PD's formulas using the
redshift-dependent matter density $\Omega_0(z)$ and cosmological 
constant $\Omega_\Lambda(z)$.

\section{Predicting the Angular Correlation Function}

As is well known, the angular correlation function $w(\theta)$
for a given population is related to the spatial correlation
function $\xi(r,t)$ by Limber's equation (Rubin 1954; Limber
1954; Groth \& Peebles 1977; Phillips et al. 1978; Peebles 1980;
Baugh \& Efstathiou 1993).
If we assume that clustering is independent of luminosity (or if
we consider an averaged correlation function for the entire
sample) and that clustering is negligible on scales compared
with the depth of the survey, then the angular correlation function
in a flat Universe is (e.g., Baugh \& Efstathiou 1993)
\begin{equation}
     w(\varpi)= { 2  \int_0^\infty \int_0^\infty \, x^4 
     a^6 p^2(x) \xi(r,t) dx du \over [\int_0^\infty\, x^2 
     a^3 p(x) dx]^2},
\label{limbereqn}
\end{equation}
where $a(t)$ is the scale factor as a function of time $t$ and
$p(x)$ is the selection function (the probability that a
source at a distance $x$ is detected in the survey).
The physical (not comoving) separation between two sources separated
by an angle $\theta$ is
\begin{equation}
     r^2=a^2(t)[u^2 + x^2 \varpi^2],
\label{separation}
\end{equation}
where $\varpi=2\sin(\theta/2)$. The total number of sources in a survey of 
solid angle $\Omega_s$ is
\begin{equation}
     N={\Omega_{s}\over{4\pi}}\int_0^\infty\, x^2 a^3 p(x) dx=
     {\Omega_{s}\over 4\pi}\int_0^\infty\, {dN \over dz} dz
\label{sourcenumber}
\end{equation}
where $dN/dz$ is the redshift distribution of sources in the
survey.

The spatial correlation function is the Fourier transform of the
spatial power spectrum $P(k,z)$:
\begin{equation}
     \xi(r,z)={1\over 2\pi^2} \int_0^\infty\, P(k,z) {\sin(kr/a)
     \over (kr/a)} k^2 dk,
\label{powerspectrum}
\end{equation}
where $k$ is the comoving wavenumber.  

Substituting (\ref{separation}), (\ref{sourcenumber}) and
(\ref{powerspectrum}) into (\ref{limbereqn}), and integrating
over $u$, one obtains
\begin{equation}
     w(\varpi)=  {1\over 2 \pi}  \int_0^\infty k\,dk \int_0^\infty \, P(k,z)
     j(k,z,\varpi)\, dz,
\label{wbande}
\end{equation}
\noindent where 
\begin {equation}
     j(k,z,\varpi)=\left({dN\over dz} {1\over N} \right)^2 {dz\over dx}
     J_0(k\varpi x).
\label {gofz}
\end {equation}

To include non-linear effects, we use (\ref{pnlk}) for the power spectrum. 
The integral in 
(\ref{wbande}) is then over $k_{NL}$ and to determine the $k_L$  
corresponding to
a given $k_{NL}$, we need $\Delta^2_{L}(k_{L})$. Before calculating the
correlation function, we therefore set up a table of
$k_{NL}$ values for a range of $k_L$'s at various redshifts.

\begin{figure}
\epsfxsize=85truemm 
\epsffile{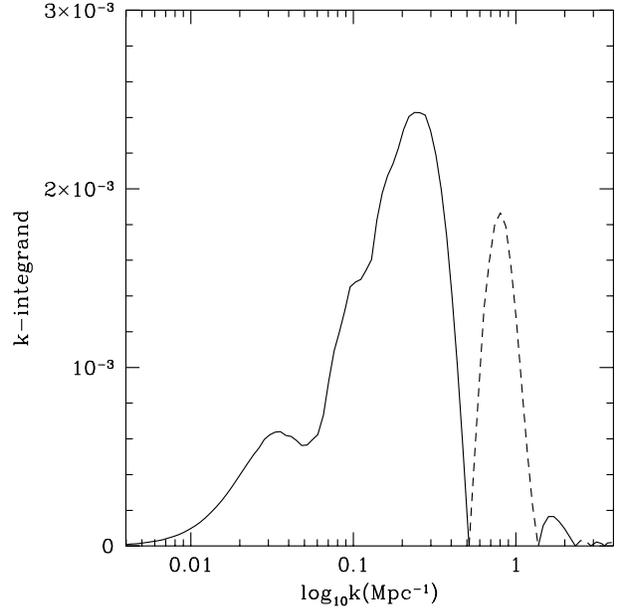}
\caption[figure3]{The $k$ integrand in (32) for $\theta=1^{\circ}$ 
(assuming $\Omega_0=1, h=0.5$ and the DP7 model for the redshift 
distribution). The dotted lines indicates negative values.}
\end{figure}

\begin{figure}
\epsfxsize=85truemm 
\epsffile{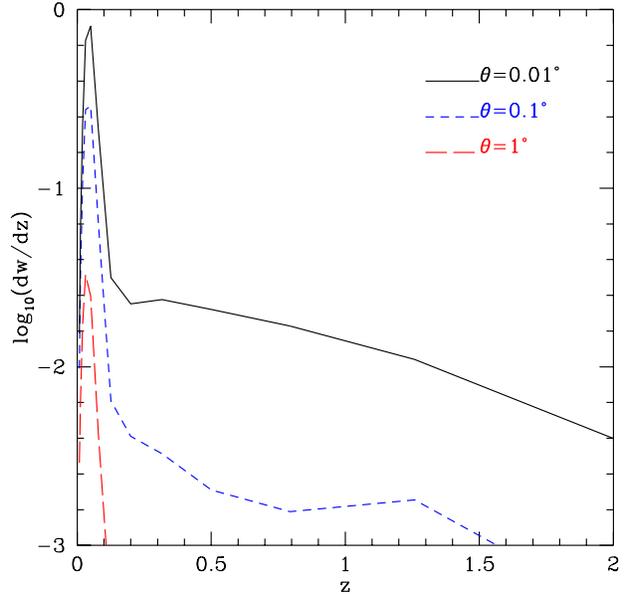}
\caption[figure4]{The distribution $dw/dz$ for a range of angles (assuming 
$\Omega_0=1, h=0.5$ and the DP7 model for the redshift distribution).}
\end{figure}

The form of the $k$ integrand in the CF estimate at $\theta=1^\circ$
is shown in Figure 2  for $\Omega_0=1$ and $h=0.5$.
One sees that there are positive contributions to the CF from $k$'s
in the range $\rm 0.01\, Mpc^{-1}$ to $\rm 0.5\, Mpc^{-1}$ with 
a large contribution coming from the larger $k$'s (smaller scales). As 
one decreases the angle $\theta$, the the peak contribution shifts
towards even larger $k$'s. 

In Figure 3 we have plotted $dw/dz$, the redshift distribution
of the clustering signal, for
a range of angles ($\Omega_0=1, h=0.5$ Universe). One sees that the 
larger-angle CF estimates 
are dominated by contributions from nearby objects and even at $0.1^\circ$
one expects about half the signal to come from objects with
$z<0.1$.  This illustrates that the angular correlation function
probes clustering at redshifts significantly smaller than those
characteristic of the survey population. 
More quantitatively, the models used here predict that
the mean redshift probed by
the correlation function at $\theta=0.01^\circ$ ($0.1^\circ$,$1^\circ$) is
expected to be $\overline z$=0.4 (0.18, 0.08). 
Together, Figures 2 and 3 highlight the fact that the CF measured
for {\sl FIRST} is likely to be rather sensitive to smallish-scale correlations in 
fairly nearby sources. 

\section {Results}
Figure 4 shows how the predictions for $w(\theta)$ vary for different 
estimates of the redshift distribution and how they change when
nonlinear evolution is considered. There appears to be a fair amount
of uncertainty in the predicted result, depending on which
redshift distribution is chosen.
It should be noted, however, that Condon's luminosity
function is more carefully constructed to fit the low-$z$ population
and it was pointed out above that these sources are expected to make large
contributions to the clustering signal. In addition, Peacock has 
indicated that DP's model 7 is the best model to use (the difference
between `high-$z$' model 7 as opposed to `medium-$z$' model 7 is not 
significant). If one limits
oneself to DP7 and Condon's model then uncertainties in the redshift
distribution do not contribute very significantly to uncertainties in
the correlation-function predictions. 

For the $\Omega_0=1, h=0.5$ model shown in Figure 4, it is clear that
the effect of nonlinear evolution is significant
on scales less than $\sim30^\prime$. For a $\Lambda$-dominated model
the nonlinear contributions are significant out to scales of
$60^\prime$.  The precise scale at which nonlinearities become
important also depends on the redshift distribution.
Note that in the standard-CDM model, $\sigma_8$ for the linear-evolution 
spectrum is 1.22 while $\sigma_8$ for the nonlinear-evolution
spectrum is 1.17.

\begin{figure}
\epsfxsize=85truemm 
\epsffile{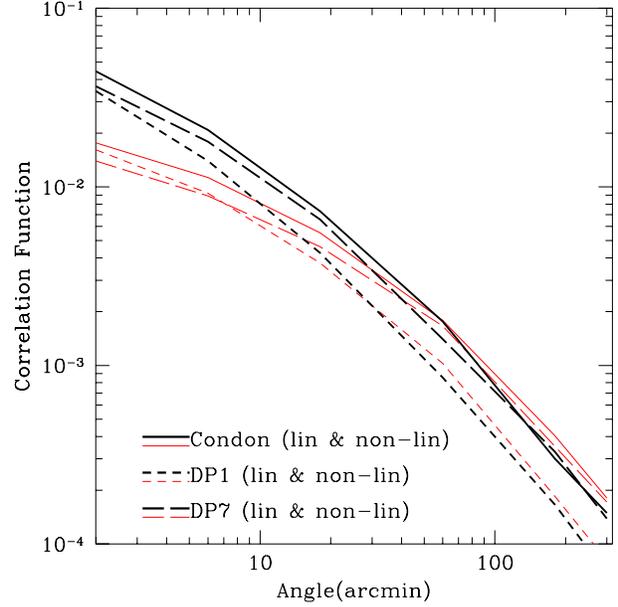}
\caption[figure5]{CDM predictions for $w(\theta)$ (assuming no bias):
the effect of varying the assumed redshift distribution and the
effect of including nonlinear clustering in an $\Omega_0=1, h=0.5$,  
Universe. Bold curves include nonlinear corrections.}
\end{figure}

\begin{figure}
\epsfxsize=85truemm 
\epsffile{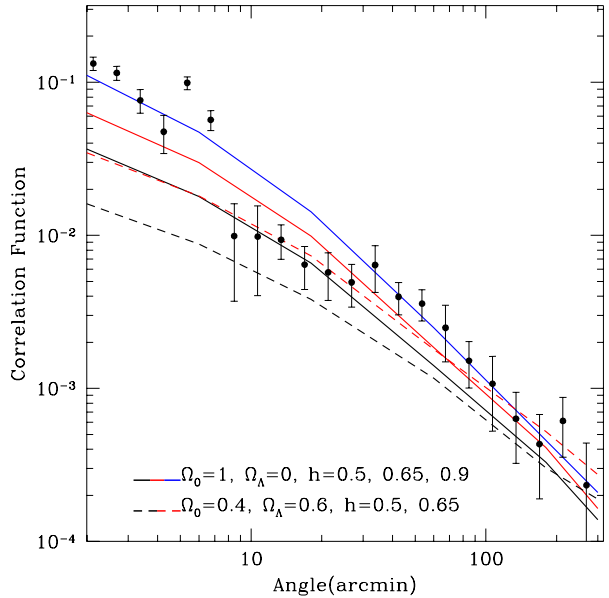}
\caption[figure6]{Comparison of the measured $w(\theta)$ with that 
predicted by various CDM models (assuming DP7 and no bias).}
\end{figure}

Figure 5 compares the measured CF in the {\sl FIRST} survey
(1-mJy flux threshold) with the CF's predicted by various cosmological 
models: the three solid lines are for $\Omega_0=1$ and $\Omega_{\Lambda}=0$
with $h$ taking on values of 0.5 (lowest curve), 0.65 and 0.9. The dashed
lines show results for $\Omega_0=0.4, \Omega_{\Lambda}=0.6$ and $h$ taking
values 0.5 and 0.65. Error bars in the plot are determined using a 
`partition bootstrap method'
in which the CF is calculated for 10 subdivisions of the survey region
and the standard deviation of these measurements at each angle is used
as a measure of the error. Note that the spike in the CF at $6^\prime$
and possibly the dip at $9^\prime$ are associated with the sidelobe
contamination discussed in Cress et al. (1996).
At a first glance it would appear that the data is best fit  
by an $\Omega_0=1$ model with a high Hubble constant, but this
is misleading. It is likely
that at low fluxes, a significant fraction of the source counts come
{}from fairly nearby ``star-bursting'' spiral galaxies as opposed to the 
AGN-powered sources that one normally associates with radio 
sources (Condon 1984). Since spiral galaxies have been shown to have 
different clustering properties
{}from ellipticals (see, for example, Hermit et al. 1996) and from AGN in 
general (for an estimate of the quasar correlation function see 
Boyle et al. 1993), it seems likely that the correlation function
measured here will have contributions from populations with different
clustering properties. In \S 3 we showed that on large
angular scales ($\theta\sim 1$), we expect the signal to be dominated
by fairly local sources and so it is probably more reflective of the
``starburst'' clustering. On small angular scales we are probably more
sensitive to the clustering of more distant AGN. With this in mind, we
do not attach special significance to the curve which best
matches the slope of the data. Instead, we
calculate the bias $b$, separately on small and
large scales (where $w_{\rm matter-matter}=b^2 w_{\rm radio-radio}$)
and use this as an indicator of the relative bias of different
populations in the survey.
Inferred biases for various cosmological models are shown 
in Table 1. 
Estimates of the bias calculated when the power spectrum is normalized to the 
observed cluster abundance are shown in brackets. These were estimated by 
$b(cluster)\approx b(COBE)\sigma_8(cluster)/\sigma_8(COBE)$ where 
$\sigma_8(COBE)$ is given by Bunn \& White (1997) and 
$\sigma_8(cluster)=0.6\Omega_0^{-0.53}$ for $\Omega_0=0.4$ (Viana
\& Liddle 1996; White, Efstathiou \& Frenk 1993).

\begin{table*}
\caption{Inferred bias, assuming DP7 for the redshift distribution}
\begin{center}
\begin{tabular}{ccccccc}
Sample & model ($\Omega_0,\Omega_\Lambda,h$) & b($2^\prime$) &b($20^\prime$) & b($100^\prime$) & $\overline z$ &$r_0$ ($h^{-1}$Mpc)\\
\hline
\hline
{\sl FIRST} 1\,mJy & 1, 0, 0.5 & 1.7(3.5) & &1.2 & 0.27 ($2^\prime$), 0.07 ($100^\prime$)&6.8\\
             & 1, 0, 0.65 & 1.3 & &1.1 & &6.8\\
             & 1, 0, 0.9 & 1 & &1 & &6.8\\
             & 0.4, 0.6, 0.65 & 1.8(2.0) & &1 & 0.3($2^\prime$)&7.1 \\
             & 0.4, 0.6, 0.5 & 2.7 & &1.3 & &7.1\\ 
WENSS 35\,mJy& 1, 0, 0.5 & &1.5(3.2) & &0.33&9.3\\
             & 0.4,0.6,0.65& & 1.5(1.7)& &0.42&10.3\\
GB/PMN 50\, mJy&1, 0, 0.5& &2.5(5.1) & &0.42& 15.7\\
             &0.4,0.6,0.65&&2.5(2.8)&&0.48&17.6\\
PN&            1,0,0.5&&1.4(2.8)&&0.05&10\\
             &0.4,0.6,0.65&&1.4(1.7)&&&10\\
\end{tabular}
\end{center}
\end{table*}

Table 1 also shows the value of $r_0$ calculated for various models
assuming that the spatial correlation function of sources is
approximated by a power law 
and that density fluctuations evolve linearly; i.e. 
$\xi(r,z)=(r/r_0)^{-\gamma}(1+z)^{-2}$ (in comoving coordinates).
We used $\gamma=2.2$ for the {\sl FIRST} measurement and $\gamma=1.8$ for the
other surveys but the $r_0$ results do not depend heavily on the chosen
value of $\gamma$. Peacock \& Nicholson (PN, 1991) measured the spatial
correlation function for bright ($\rm S_{2.4GHz}>1Jy$) radio galaxies
with redshifts between 0.01 and 0.1 and found it to be well fit
by $r_0=11\,h^{-1}$ Mpc and $\gamma=1.8$.  After correcting for 
redshift-space distortions, one would most likely obtain a real-space 
$r_0$ slightly less than this.
This result indicates that bright radio galaxies trace the matter distribution
in a way similar to low-richness Abell clusters. 
Assuming linear evolution, estimates for $r_0$ in the {\sl FIRST} sample
range from $6h^{-1}\rm Mpc$ to $8h^{-1}\rm Mpc$ depending on what redshift 
distribution is used. Note that in
Cress et al. (1996), the $r_0$ obtained was based on a different estimate
of the angular correlation function. The value obtained here is
consistent with the fainter sources in the sample being less clustered
than the bright sources investigated by PN. This picture is supported 
by Ledlow \& Owen's (1995) observation that in a sensitive radio survey,
field ellipticals and cluster ellipticals have a similar probability of
being detected. One might thus expect
sources to be clustered more like ellipticals than like clusters.

Another consistency check is supplied by the measurement of the angular 
cross-correlation of Abell clusters
and {\sl FIRST} sources presented in Cress et al. (1996). Using
this result in a cross-correlation analog of Limber's equation
(see Seldner \& Peebles 1978),
we inferred the amplitude of the spatial cross-correlation of
Abell clusters and {\sl FIRST} sources.  DP's model 7 was used for the redshift
distribution of {\sl FIRST} sources and the distribution of Abell clusters 
was obtained by fitting a function of the form $\log N(z)=-9.1z-3.5$
to data given in Huchra et al. (1990). By combining the 
autocorrelation and the cross-correlation we found the 
ratio of cluster bias to 
radio-source bias ($b_c/b_r$) to be about 1.9. A similar estimate of
$b_c/b_r$ was also 
obtained by combining the measured cluster-cluster spatial 
correlation and the radio-radio spatial correlation function inferred above. 
Mo, Peacock \& Xia (1993) compared bright radio galaxies with
Abell clusters and obtained a value of 1.7 for this ratio. This is 
also consistent with the idea that the faint sources in {\sl FIRST}
are somewhat less clustered than the bright sources investigated
by their work (same as in PN).

\begin{figure}
\epsfxsize=85truemm 
\epsffile{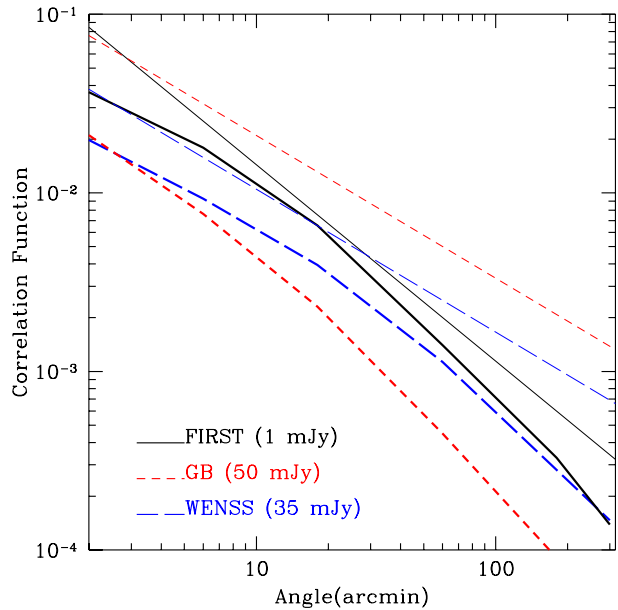}
\caption[figure7]{Comparison of the measured $w(\theta)$ with that 
predicted by standard CDM (assuming no bias) for 3 different surveys.
DP7 was used to model the redshift distribution. Power law fits to 
the data are shown as faint lines.}
\end{figure}

Angular correlation functions of radio sources have also
been measured for the Green Bank and the Parkes-MIT-NRAO (PMN)
surveys done at 4.85 GHz (Loan, Wall \& Lahav 1997; 
Kooiman et al. 1995; Sicotte 1995)
and for WENSS --- the 325-MHz Westerbork Northern 
Sky Survey (Rengelink et al. 1997). We have investigated CDM predictions
for these surveys using DP7 to estimate the redshift
distribution of sources with $S_{4.85}> 50$ mJy in the GB/PMN surveys 
and of sources with $S_{325}> 35$ mJy in WENSS. Predicted correlation
functions (assuming no bias) are shown in Figure 6.   

Taking $A_{\rm WENSS}$=0.0025, $A_{\rm GB/PMN}=0.005$ and $\gamma=1.8$ 
[where $w(\theta)=A \theta^{1-\gamma}$], we have calculated the bias
of sources in these surveys relative to CDM predictions and listed them
in Table 1. 
The source composition in these two surveys
is probably more similar to the sample investigated by PN, although
WENSS is likely to include fewer quasars. The biases inferred for 
the PN sample are also shown in the Table.

While a large amount of variation in inferred clustering strength could be
attributed to uncertainties in the correlation-function estimates and
to the different clustering of different populations,
data summarized in Table 1 show some
indication that as one probes deeper, the inferred bias
increases. Assuming that {\sl FIRST} (at small angles) and the other surveys
probe a similar population of objects, we have, with cluster normalization, 
$b_{radio AGN}\sim (2.8,\,3.4,\,3.1,\,5.1)$
at ${\overline z}\sim (0.05,\,0.27,\,0.33,\,0.42)$ for the standard model
and $b_{radio AGN}\sim (1.7,\,2.0,\,1.7,\,2.8)$ at 
${\overline z}\sim (0.05,\,0.30,\,0.42,\,0.48)$.
One interpretation of this is that the clustering is
evolving slower than CDM models predict. However, Matarrese et al. (1996), 
drawing on work by Mo \& White (1996) and by Fry (1996), have recently 
argued that in CDM models, the bias of a given population {\it
should} decrease with time; 
that is, the fluctuations in density inferred from luminous objects
should tend towards the real fluctuations in the mass density as time
goes by. Thus, the observations in Table 1 could still be consistent with 
CDM predictions if bias evolution takes place. Interestingly enough,
this apparent bias evolution is stronger than that seen in
optical surveys (Matarrese et al. 1996).  This would be consistent
with the general picture that the bias evolution for objects
which tend to form at higher-density mass peaks is stronger. 
If one writes $b(z)=1+b_0(1+z)^\beta$, the value of $\beta$ which best
fits the data is $\sim 2$ but this can vary significantly when uncertainties
in the measurements are considered.

\section {Conclusions}
We have studied the implications of angular clustering of radio
sources in deep surveys for standard-CDM--like models for
structure formation.  We have examined the effect of
uncertainties in the redshift distribution and considered
nonlinear evolution of the power spectrum.  We have also studied
what may be learned about the biasing of such sources and the
evolution of such a bias.

We have shown that uncertainties in the redshift distribution of
{\sl FIRST} sources can contribute a significant amount to the uncertainties 
in the CDM predictions of the angular correlation function 
even when bias is not considered.  However, limiting oneself to
Condon's model and DP7 reduces the uncertainties significantly.
It should also be kept in mind that the uncertainty in the
predicted angular correlation function which arise from
uncertainties in the redshift distribution are still not much larger
than the statistical uncertainty in the measured angular
correlation function.  Therefore, uncertainties in the redshift
distribution do not currently affect strongly the implications
of radio-source clustering for the origin of structure.

The effect of nonlinear evolution on the predictions has also been 
explored. It appears that nonlinear contributions become important
on scales of $10^\prime$ to $60^\prime$ depending on what model 
is used. Therefore, as data improve, the standard approximation
of a power-law evolution of the clustering with redshift will
break down.  We also found that although the mean redshift of
sources in deep radio surveys may be of order unity, the
clustering signal comes primarily from smaller redshifts.
That is, the angular correlation function is more sensitive to
nearby clustering than one might expect from the $dN/dz$
distribution.  Furthermore, the redshift distribution of the
clustering signal varies significantly with the angular scale
probed.  Therefore, the angular correlation function reflects
the evolution of the three-dimensional power spectrum as well as
the power spectrum at some fixed epoch.

We have shown that spatially-flat CDM models can produce angular
correlation functions similar to those observed in the {\sl FIRST} survey
as long as some kind of biasing is invoked.  The needed biasing
seems to be a bit larger than that needed for optical sources,
but less than that needed for brighter radio sources.  This is
probably due to the fact that {\sl FIRST} sources are a heterogeneous
sample which contain numerous nearby starbursting galaxies as
well as more distant brighter radio sources.  We have also
begun to explore what may be learned about the evolution of
bias from the {\sl FIRST} and other radio surveys.  At this stage, it
is difficult to make very quantitative statements about the
evolution of clustering because the CF measurements that probe
higher-$z$ ranges have large uncertainties associated with them.
In addition, bias evolution 
will have to be understood better if we are to learn about the
evolution of mass clustering.  This point is particularly
significant for a heterogeneous sample of objects such as {\sl FIRST}
sources.

One way to improve our ability to 
make quantitative statements about clustering evolution
is to make more precise measurements of the clustering of sources that
probe the higher-$z$ range. As the {\sl FIRST}
survey coverage increases, we will be able to obtain significant
clustering signals for samples with higher flux thresholds.
Increasing the flux threshold should increase the average redshift 
probed by the correlation function and thus improves one's
chances of probing larger scales. Another way of increasing the
redshift probed by the CF might be to remove all sources that
have optical counterparts in the APM survey. This is currently
being investigated. Improving our estimates of the redshift 
distribution of faint sources 
is also important and there is some hope that this will be done in
the near future.

\section*{Acknowledgments}

We thank John Peacock, Jim Condon, Ofer Lahav, David Helfand,
Alexandre Refregier and Jacqueline van Gorkom for useful 
comments. This work was supported by the U. S. Department of
Energy under contract DE-FG02-92ER40699, NASA grant NAG5-3091
and the Alfred P. Sloan Foundation. The FIRST project is supported
by grants from the National Geographic Society, 
NSF grant AST94-19906, NATO, IGPP, Columbia University, 
and Sun Microsystems.


\end{document}